\documentclass[lettersize,journal]{IEEEtran}
\usepackage{amsmath,amsfonts}
\usepackage{algorithmic}
\usepackage{epsfig, algorithm}
\usepackage{array}
\usepackage{amssymb}
\usepackage{acronym} 
\usepackage[caption=false,font=normalsize,labelfont=sf,textfont=sf]{subfig}
\usepackage{textcomp}
\usepackage{stfloats}
\usepackage{tabularx}
\usepackage{xcolor}
\usepackage{url}
\usepackage{verbatim}
\usepackage{multirow}
\usepackage{acronym}
\usepackage{graphicx}	
\usepackage{stfloats}
\usepackage{soul}
\usepackage{cancel}
\usepackage{cite}
\def\BibTeX{{\rm B\kern-.05em{\sc i\kern-.025em b}\kern-.08em
T\kern-.1667em\lower.7ex\hbox{E}\kern-.125emX}}
\usepackage{balance}
\usepackage{multirow}
\usepackage{hhline}
\input{AcronymsListFinal}
\acresetall
\usepackage{fancyhdr}

\begin{document}
\title{Unified Interference-Aware Water-Filling for QoS-Constrained Communication, Sensing, and JRC}
%%%%%%%%%%%%%%%%%%%%%%%%%%%
\author{Ahmed~Naeem, Anastassia~Gharib  \IEEEmembership{Member,~IEEE}, Hüseyin~Arslan~\IEEEmembership{Fellow,~IEEE}
\\This work has been submitted to the IEEE for possible publication. Copyright may be transferred without notice, after which this version may no longer be accessible.}
\markboth{Journal of \LaTeX\ Class Files,~Vol.~14, No.~8, June~2025}%
{Shell \MakeLowercase{\textit{et al.}}: Bare Demo of IEEEtran.cls for IEEE Journals}
%%%%%%%%%%%%%%%%%%%%%%%%%%%
\maketitle 
%%%%%%%%%%%%%%%%%%%%%%%%%%%
\begin{abstract}
Water-filling (WF) algorithms are pivotal in maximizing capacity and spectral efficiency in multiple-input and multiple-output (MIMO) systems. However, traditional WF approaches cater solely to communication requirements, neglecting the emerging heterogeneity of 6G, including sensing and joint radar-communication (JRC). As these diverse demands grow in importance and have different Quality of Service (QoS) constraints, traditional WF becomes inadequate. Therefore, in this paper, we propose a unified interference-aware and QoS-constrained  WF algorithm for systems with communication, sensing, and JRC. The proposed algorithm enables power allocation for multi-user MIMO systems, effectively addressing interference and balancing the support for heterogeneous user requirements.
\end{abstract}
%%%%%%%%%%%%%%%%%%%%%%%%%%%
\section{Introduction}
\par As networks evolve toward 6G, \ac{UE} needs now encompass communication, sensing, and \ac{JRC}, demanding more advanced algorithms \cite{10458884}. This is because in 6G, \acp{UE} are not only mobile phones requiring the communication service, but also \ac{IoT} devices requiring the sensing service \cite{9990579}. A temperature sensor in a smart city, for instance, requires only sensing service for precise localization, while it utilizes other communication technologies to exchange data with other sensors. In contrast, an autonomous vehicle needs both 6G communication and sensing services. Still, efficient \ac{PA} remains vital in 6G networks to maximize sum capacity, enhance data rates, and maintain reliability under dynamic channel conditions and increasing \ac{UE} demands. 
\par \Ac{WF} is a fundamental technique in \ac{MIMO} systems for optimizing power across sub-channels based on channel gains. Existing \ac{WF} algorithms cannot address these heterogeneous demands due to their communication-centric design \cite{10436227, 10989634, 10908220}. For example, the work in \cite{yu2007multiuser} proposes a modified algorithm by solving a nonconvex \ac{MU} rate maximization problem under crosstalk. In \cite{popescu2007simultaneous}, a simultaneous \ac{WF} approach is introduced for two mutually interfering transceivers using non-cooperative coding. Authors in \cite{viswanath2002asymptotically} develop an asymptotically optimal \ac{WF} for vector multiple-access channels to maximize sum capacity by analyzing its performance across \ac{UE} loads and \ac{SNR}. They further extend the proposed algorithm to \ac{MU} classes with power constraints. In \cite{wang2015iterative}, an iterative dynamic \ac{WF} algorithm is developed for energy scheduling in an \ac{MU} fading multiple-access channel with energy harvesting. A low-complexity \ac{DL} training design using \ac{WF} is proposed to maximize sum rate under limited coherence time in \cite{naser2023downlink}. Equal \ac{PA} via \ac{WF} is addressed in \cite{chauhan2023power} by solving the maximization problem with Lagrange multipliers.
\par Although the aforementioned \ac{WF} algorithms are effective, they lack support for heterogeneous \ac{UE} requirements of communication, sensing, and \ac{JRC}. Hence, advanced \ac{WF} is required to meet the need of these \ac{QoS} requirements, ensuring fairness for each \ac{UE}. Thus, this paper proposes a unified \ac{WF} algorithm that addresses interference, \ac{QoS} constraints, and service prioritization, offering an efficient solution for future 6G networks. To overcome traditional \ac{WF} limitations, this work makes the following key contributions.
\begin{itemize}
    \item We formulate a unified, interference-aware \ac{WF} optimization that jointly supports heterogeneous \ac{UE} requirements through customized utility functions and \ac{QoS} constraints.
    \item We derive closed-form power update formulas via \ac{KKT} conditions with a fairness-driven subgradient method to ensure proportional \ac{PA}.
    \item We develop an iterative approximation for the proposed \ac{WF} optimization problem to reduce system complexity.
\end{itemize}
We evaluate the proposed \ac{WF} algorithm over traditional and equal-power algorithms across varying system conditions.
%%%%%%%%%%%%%%%%%%%%%%%%%%%%%%%%%%%%%%%%%%%%%%%%%%
\section{System Model} \label{sec:system_model}
\par As shown in Fig.~\ref{FigMAIN1}, we consider a single-cell system, where a \ac{BS} with a \ac{UPA} of \(N_t\) antennas serves \(K\) \acp{UE}. These \acp{UE} are heterogeneous, representing mobile phones, autonomous vehicles, and other \ac{IoT} devices. Therefore, they have different requirements when it comes to communication and sensing. We assume that each \emph{k-th} \ac{UE} has \(N_{r,k}\) antennas and is categorized into one of the three disjoint sets based on their service requirements: communication \((\mathcal{K}_C)\), sensing \((\mathcal{K}_S)\), and \ac{JRC} \((\mathcal{K}_{JRC})\), where $|\mathcal{K}_C| + |\mathcal{K}_S| + |\mathcal{K}_{JRC}| = K$. 
\par The channel to the \emph{k-th} \ac{UE} (\(\mathbf{H}_k \in \mathbb{C}^{N_{r,k} \times N_t}\)) is assumed constant over a coherence interval \(\tau_{c,k}\). Each \ac{UE} type has a distinct coherence time, corresponding to the duration of its channel being stationary. To ensure fairness and synchronization between \acp{UE}, the system is updated every $\tau_c = \min_{k=1,\ldots,K} (\tau_{c,k}),$ corresponding to the shortest coherence time among all \acp{UE}. We assume that \ac{PA} updates are done every \(\tau_c\) for simultaneous updates of heterogeneous \acp{UE}.
\par Based on the \emph{k-th} \ac{UE} requirement, the transmit signal~is:
%%%%%%%%%%%%%%%%%%%%%%%
\begin{equation} \small
\mathbf{x}_k = 
\begin{cases} 
\sqrt{P_{C,k}}~\mathbf{W}_{C,k}~\mathbf{s}_{C,k}, & \text{if UE } k \in \mathcal{K}_C \\
\sqrt{P_{S,k}}~\mathbf{W}_{S,k}~\mathbf{s}_{S,k}, & \text{if UE } k \in \mathcal{K}_S \\
\sqrt{P_{JRC,k}}~\mathbf{W}_{JRC,k}~\mathbf{s}_{JRC,k}, & \text{if UE } k \in \mathcal{K}_{JRC}
\end{cases},
\label{eq:tx_signal}
\end{equation}
%%%%%%%%%%%%%%%%%%%%%%%
where \(\mathbf{s}_{C,k} \in \mathbb{C}^{N_{r,k}}\), \(\mathbf{s}_{S,k} \in \mathbb{C}^{N_{r,k}}\), and  \(\mathbf{s}_{JRC,k} \in \mathbb{C}^{N_{r,k}}\) are the communication, sensing, and \ac{JRC} \ac{DFRC} signals, respectively. The corresponding beamforming matrices and \ac{PA} for sub-channel $i$ are \(\mathbf{W}_{C,k} \in \mathbb{C}^{N_t \times N_{r,k}}\), \(\mathbf{W}_{S,k} \in \mathbb{C}^{N_t \times N_{r,k}}\),  \(\mathbf{W}_{JRC,k} \in \mathbb{C}^{N_t \times N_{r,k}}\) and \(P_{C,k,i}\), \(P_{S,k,i}\), \(P_{JRC,k,i}\), respectively. The received signal at the \emph{k-th} \ac{UE} is therefore given by $\mathbf{y}_k = \sqrt{P_{k,i}} \mathbf{H}_k \mathbf{W}_k \mathbf{s}_k + \sum_{j \neq k} \sqrt{P_{j,i}} \mathbf{H}_k \mathbf{W}_j \mathbf{s}_j + \mathbf{n}_k + \mathbf{c}_k,$ where \(P_{k,i}\), \(\mathbf{W}_k\), and \(\mathbf{s}_k\) are selected from the corresponding $\{P_{C,k,i}, P_{S,k,i}, P_{JRC,k,i}\}$, \(\{\mathbf{W}_{C,k}, \mathbf{W}_{S,k}, \mathbf{W}_{JRC,k}\}\), and \(\{\mathbf{s}_{C,k}, \mathbf{s}_{S,k}, \mathbf{s}_{JRC,k}\}\), respectively. Furthermore, \(\mathbf{n}_k \sim \mathcal{CN}(0, N_{0,k} \mathbf{I}_{N_{r,k}})\) is the additive noise, and \(\mathbf{c}_k \sim \mathcal{CN}(0, C_{0,k} \mathbf{I}_{N_{r,k}})\) is the clutter.
%%%%%%%%%%%%%%%%%%%%%%%%%%%%%%%%%%%%%%%%%%%%%%
\section{Unified Water-filling Problem Formulation}
\label{sec:proposed_wf}
\par Each \ac{UE} type has specific performance metrics, comprising the desired signal, noise, and \ac{MU} interference. For \ac{UE} \(k \in \mathcal{K}_C\), capacity maximization is important using \ac{ZF} precoding to suppress \ac{MU} interference. The channel matrix is decomposed via singular value decomposition \(\mathbf{H}_k = \mathbf{U}_k \mathbf{\Sigma}_k \mathbf{V}_k^H\), with \(\mathbf{\Sigma}_k = \text{diag}(\sqrt{\lambda_{k,1}}, \ldots, \sqrt{\lambda_{k,r_k}})\). Even though the \ac{ZF} precoder \(\mathbf{W}_{C,k}\) is designed to nullify interference to other \acp{UE} (\(\mathbf{H}_j \mathbf{W}_{C,k} \approx \mathbf{0}, j \neq k\)), some residual interference remains. Hence, the resulting capacity is:
%%%%%%%%%%%%%%%%%%%%%%%%%%%%%%%%%%%%%%%%%%%%%%%%%%
\begin{figure}
\centering 
\resizebox{0.8\columnwidth}{!}{
\includegraphics{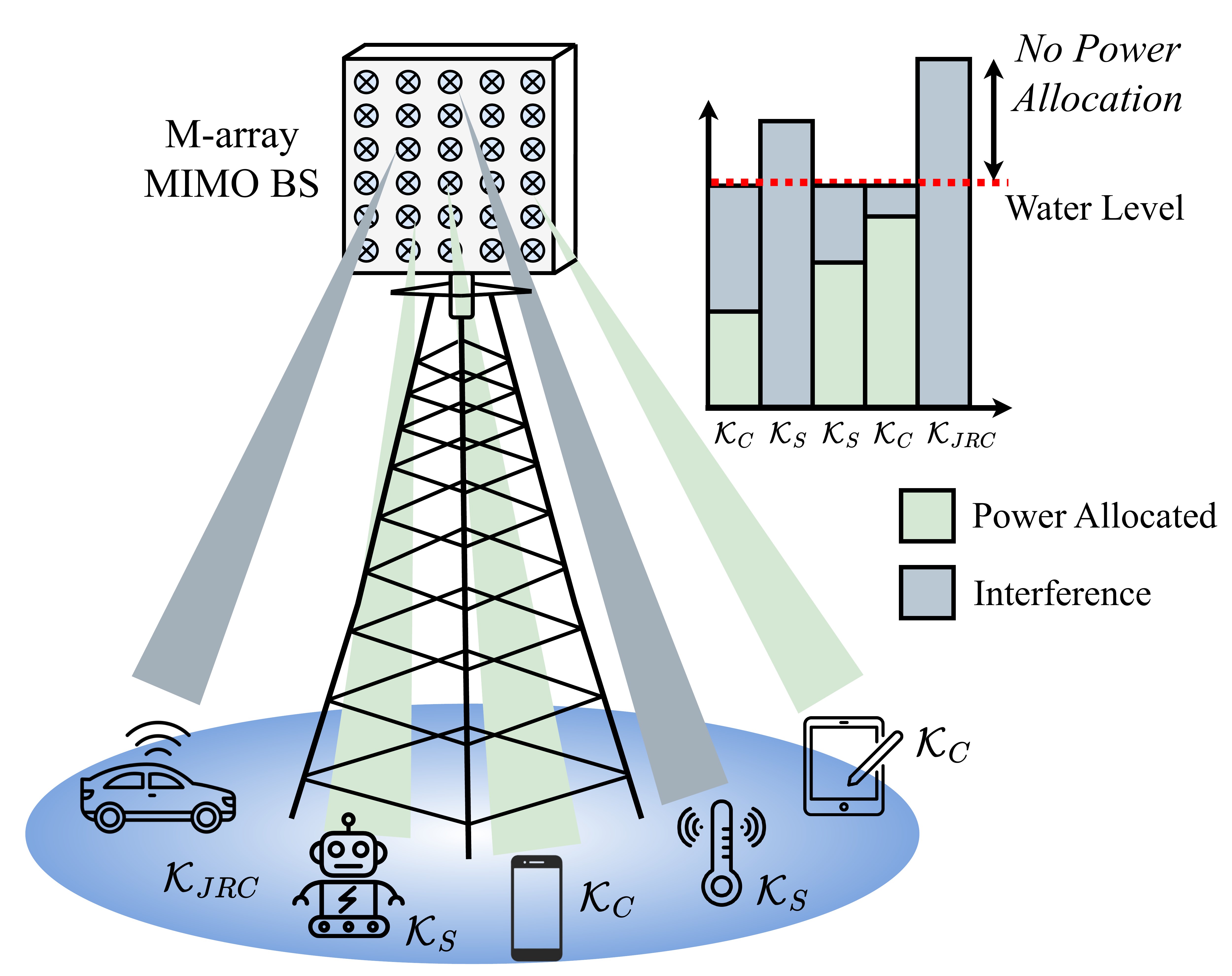}}
\caption{Proposed \ac{MU} \ac{MIMO} system architecture.}
\label{FigMAIN1}
\end{figure}
%%%%%%%%%%%%%%%%%%%%%%%%%%%%%%%%%%%%%%%%%%%%%%%%%%
\begin{equation} \small
C_k = \sum_{i=1}^{r_k} B \log_2 \left( 1 + \frac{P_{C,k,i} \lambda_{k,i}}{N_{0,k} + \sum_{j \neq k} P_{j,i} |\mathbf{H}_k \mathbf{w}_{j,i}|^2} \right),
\label{eq:capacity}
\end{equation}
%%%%%%%%%%%%%%%%%%%%%%%
where \(r_k = \min(N_t, N_r)\) is channel rank, \(B\) is bandwidth, and \(\mathbf{w}_{j,i}\) is the \(i\)-th column of precoder \(\mathbf{W}_j\).
For \ac{UE} \(k\in\mathcal{K}_S\), the performance is measured by \ac{SCNR} to quantify target detection against clutter. Using matched precoding (\(\mathbf{W}_{S,k} = \mathbf{V}_k\)), the SCNR is:
%%%%%%%%%%%%%%%%%%%%%%%
\begin{equation} \small
\text{SCNR}_k = \frac{\sum_{i=1}^{r_k} P_{S,k,i} \lambda_{k,i}}{N_{0,k} + C_{0,k} + \sum_{j \neq k} P_{j,i} |\mathbf{H}_k \mathbf{w}_{j,i}|^2}.
\label{eq:scnr}
\end{equation}
%%%%%%%%%%%%%%%%%%%%%%%
For \ac{UE} \(k \in \mathcal{K}_{JRC}\), \ac{ZF} precoding is applied for communication and matched precoding for sensing. Therefore, both objectives are considered as a weighted function as follows:
%%%%%%%%%%%%%%%%%%%%%%%
\begin{equation} \small
\begin{aligned}
U_k &= \alpha_k \sum_{i=1}^{r_k} B \log_2 \left( 1 + \frac{P_{JRC,k,i} \lambda_{k,i}}{N_{0,k} + \sum_{j \neq k} P_{j,i} |\mathbf{H}_k \mathbf{w}_{j,i}|^2} \right) \\
&\quad + (1 - \alpha_k) \frac{\sum_{i=1}^{r_k} P_{JRC,k,i} \lambda_{k,i}}{N_{0,k} + C_{0,k} + \sum_{j \neq k} P_{j,i} |\mathbf{H}_k \mathbf{w}_{j,i}|^2},
\end{aligned}
\label{eq:jrc_utility}
\end{equation}
where \(\alpha_k \in (0,1)\) is a weighting parameter, balancing communication and sensing priorities.
%%%%%%%%%%%%%%%%%%%%%%%
\par Since our goal is to simultaneously optimize \ac{PA} for heterogeneous \acp{UE}, an optimization problem is formulated to maximize a weighted utility function that balances \(C_k\) and \(\text{SCNR}_k\) for each \ac{UE} $k$ as follows:
%%%%%%%%%%%%%%%%%%%%%%%
\begin{equation} \small
\begin{aligned}
& \max_{P_{k,i}} \sum_{k \in \mathcal{K}} \sum_{i=1}^{r_k} \alpha_{k,i} \beta_{k,i} C_{k,i} + (1 - \alpha_{k,i}) \beta_{k,i} \text{SCNR}_{k,i} \\
& \text{subject to:} \\
& \sum_{k \in \mathcal{K}} \sum_{i=1}^{r_k}{P_{k,i}} \beta_{k,i} \leq P_{\text{total}}, \\
& P_{k,i}\beta_{k,i} \geq 0, \quad \forall~ \text{UE}~k \in \{\mathcal{K}_C \cup \mathcal{K}_S \cup \mathcal{K}_{JRC}\}, \, \forall~i, \\
 & C_k\beta_{k,i} \geq \alpha_{k,i}C_{k,i,\text{min}}, \quad \forall~ \text{UE}~k \in \{\mathcal{K}_C \cup \mathcal{K}_{JRC}\}, \\
& \text{SCNR}_k\beta_{k,i} \geq (1 - \alpha_{k,i})S_{k,i,\text{min}}, ~ \forall~ \text{UE}~k \in \{\mathcal{K}_S \cup \mathcal{K}_{JRC}\},
\end{aligned}
\label{main}
\end{equation}
%%%%%%%%%%%%%%%%%%%%%%%
where $\alpha_{k,i} \in [0,1]$, taking zero value for \ac{UE} $k \in \mathcal{K}_S$, unity value for \ac{UE} $k \in \mathcal{K}_C$, and in between ($\alpha_{k,i}=\alpha_k$) for \ac{UE} $k \in \mathcal{K}_{JRC}$. A binary channel quality factor \(\beta_{k,i}\) is used to eliminate the low-gain sub-channels from allocating power to them, eliminating unnecessary interference. The problem in \eqref{main} incorporates fairness through the logarithmic utility functions implicit in \(C_k\) and \(\text{SCNR}_k\), ensuring equitable performance across all UE types while satisfying QoS constraints (\(C_{k,i,\text{min}}\) and \(S_{k,i,\text{min}}\)). By accounting for interference, noise, and clutter, and updating every \(\tau_c\), the algorithm delivers a robust \ac{PA} strategy for heterogeneous \acp{UE}. However, \eqref{main} is non-convex due to the interference-dependent terms in \(C_k\) and \(\text{SCNR}_k\), which couple the \ac{PA} across \acp{UE}. This complexity is compounded by the joint constraints for communication and sensing \ac{QoS}. In the next section, we propose a solution to this~problem.
% To address this, we next propose an iterative algorithm that approximates this non-convex problem by treating interference as constant within each iteration.
%%%%%%%%%%%%%%%%%%%%%%%%%%%%%%%%%%%%%%%%%%%%%%%%%%%%%%
\section{The Proposed Iterative Algorithm}
\label{sec:proposed_solution}
To address the non-convexity of \eqref{main}, we propose an iterative unified \ac{WF} algorithm shown in Algorithm \ref{alg:WF}. It iteratively computes \ac{PA} values for each \ac{UE} type and adjusts dual variables to satisfy \ac{QoS} and fairness until convergence is achieved. To begin with, we initialize \ac{PA} to each \ac{UE} as zero. After this, the iterative process begins. For each \emph{k-th} \ac{UE}, we define \( I_{k,i} \) as a constant to be treated as an interference term. This interference term is assigned to \( \sum_{j \neq k} P_{j,i} |\mathbf{H}_k \mathbf{w}_{j,i}|^2 \) from \eqref{eq:capacity}, \eqref{eq:scnr}, and \eqref{eq:jrc_utility} for communication, sensing, and \ac{JRC} \acp{UE}, respectively. Such approximation transforms \eqref{main} into a set of convex subproblems, enabling efficient optimization via \ac{KKT} conditions while satisfying fairness and \ac{QoS} constraints.
%%%%%%%%%%%%%%%%%%%%%%%
%%%%%%%%%%%%%%%%%%%%%%%%%%%%%%%%%%%%%%%%%%%%%%%%%%
\begin{algorithm}
\caption{Proposed Iterative Water-Filling Algorithm}
\begin{algorithmic}[1]
\STATE \textbf{Input:} $\lambda_{k,i}, C_{0,k}, \alpha_{k,i}, \beta_{k,i}, C_{k,i,\min}, S_{k,i,\min}$, $P_{\text{total}}, \tau_c$
\STATE \textbf{Initialize:} $P_{k,i}^{(0)} \gets \frac{P_{\text{total}}}{\sum_{k} r_k}$, $t \gets 0$
\REPEAT
    \FORALL{$k \in \mathcal{K}, i = 1,\dots,r_k$}
        \STATE $I_{k,i}^{(t)} \gets \sum_{j \neq k} P_{j,i}^{(t)} |\mathbf{H}_k \mathbf{w}_{j,i}|^2$
    \ENDFOR
    \STATE Compute $P_{k,i}^{(t+1)}$:
    \STATE $P_{C,k,i}$ via \eqref{eq:power_comm}, $P_{S,k,i}$ via \eqref{eq:power_sense}, and $P_{JRC,k,i}$ via \eqref{eq:power_jrc}
    \STATE Update Lagrange multipliers: $\mu$ via bisection search, and $\nu_{k,i}, \eta_{k,i}$ via subgradient updates in \eqref{eq:subgradient_update}
    \STATE $t \gets t + 1$
\UNTIL{$\| \mathbf{P}^{(t)} - \mathbf{P}^{(t-1)} \|_2 < \epsilon$}
\STATE \textbf{Output:} $\mathbf{P}=\{P_{k,i}\}~{\forall k, i}$ for the next $\tau_c$
\end{algorithmic}
\label{alg:WF}
\end{algorithm}
%%%%%%%%%%%%%%%%
\subsection{Derivations of KKT-Based Conditions}
A convex subproblem is solved using the Lagrangian method and \ac{KKT} conditions. The Lagrangian is denoted as:
%%%%%%%%%%%%%%%%%%%%%%%
\begin{equation} \small
\begin{aligned}
&\mathcal{L} = \sum_{k \in \mathcal{K}} \left[ \alpha_{k,i} \beta_{k,i} C_k + (1 - \alpha_{k,i}) \beta_{k,i} \text{SCNR}_k \right]\\& - \mu \left( \sum_{k \in \mathcal{K}} P_{k,i} \beta_{k,i} - P_{\text{total}} \right) - \sum_{k \in \mathcal{K}_C \cup \mathcal{K}_{JRC}} \nu_{k,i} (C_{k,i,\text{min}} - C_k \beta_{k,i}) \\
& - \sum_{k \in \mathcal{K}_S \cup \mathcal{K}_{JRC}} \eta_{k,i} (S_{k,i,\text{min}} - \text{SCNR}_k \beta_{k,i})  + \sum_{k \in \mathcal{K}} \sum_{i=1}^{r_k} \gamma_{k,i} P_{k,i} \beta_{k,i},
\end{aligned}
\label{eq:lagrangian}
\end{equation}
%%%%%%%%%%%%%%%%%%%%%%%
where \( \mu \geq 0 \), \( \nu_{k,i} \geq 0 \), \( \eta_{k,i} \geq 0 \), and \( \gamma_{k,i} \geq 0 \) are Lagrange multipliers for the power, communication QoS, sensing QoS, and non-negativity constraints, respectively. Taking partial derivative of \( \mathcal{L} \) to \( P_{k,i} \) and setting it to zero, we derive the \ac{PA}. We consider cases where \( \gamma_{k,i} = 0 \) (positive power) and \( \gamma_{k,i} > 0 \) (zero power).
%%%%%%%%%%%%%%%%%%%%%%%%%%%
For the communication \acp{UE}, the objective is to maximize \eqref{eq:capacity}. Therefore, the derivative is as:
%%%%%%%%%%%%%%%%%%%%%%%
\begin{equation} \label{eq:kkt_comm} \small
\begin{aligned}
&\frac{\partial \mathcal{L}}{\partial P_{C,k,i}} = \beta_{k,i} \frac{B}{\ln 2} \frac{\lambda_{k,i}}{N_{0,k} + I_{k,i} + P_{C,k,i} \lambda_{k,i}} + \nu_{k,i} \beta_{k,i} \frac{B}{\ln 2} \\&\frac{\lambda_{k,i}}{N_{0,k} + I_{k,i} + P_{C,k,i} \lambda_{k,i}} - \mu \beta_{k,i} + \gamma_{k,i} \beta_{k,i} = 0.
\end{aligned}
\end{equation}
%%%%%%%%%%%%%%%%%%%%%%%%%%%
If \( P_{C,k,i} > 0 \) (\( \gamma_{k,i} = 0 \)), then:
%%%%%%%%%%%%%%%%%%%%%%%%%%%
\begin{equation}\small
P_{C,k,i} = \left[ \frac{B (1 + \nu_{k,i})}{\mu \ln 2} - \frac{N_{0,k} + I_{k,i}}{\lambda_{k,i}} \right]^+,
\label{eq:power_comm}
\end{equation}
%%%%%%%%%%%%%%%%%%%%%%%
where \( [x]^+ = \max(0, x) \). If \( P_{C,k,i} = 0 \), then \( \gamma_{k,i} > 0 \), and $\gamma_{k,i} = \frac{B (1 + \nu_{k,i})}{\ln 2} \frac{\lambda_{k,i}}{N_{0,k} + I_{k,i}} - \mu,$ ensuring \ac{KKT} condition holds.
%%%%%%%%%%%%%%%%%%%%%%%%%%%
\par For the sensing \acp{UE}, the objective is to maximize \eqref{eq:scnr}. The derivative is defined as:
%%%%%%%%%%%%%%%%%%%%%%%
\begin{equation} \label{eq:kkt_sense}\small
\begin{aligned}
&\frac{\partial \mathcal{L}}{\partial P_{S,k,i}} = \beta_{k,i} \frac{\lambda_{k,i}}{N_{0,k} + C_{0,k} + I_{k,i}} + \eta_{k,i} \beta_{k,i} \\& \frac{\lambda_{k,i}}{N_{0,k} + C_{0,k} + I_{k,i}} - \mu \beta_{k,i} + \gamma_{k,i} \beta_{k,i} = 0.
\end{aligned}
\end{equation}
%%%%%%%%%%%%%%%%%%%%%%%
When \( P_{S,k,i} > 0 \) (\( \gamma_{k,i} = 0 \)), then the \ac{PA} to the sensing \acp{UE} is as follows:
%%%%%%%%%%%%%%%%%%%%%%%
\begin{equation}\small
P_{S,k,i} = \left[ \frac{(1 + \eta_{k,i}) \lambda_{k,i}}{\mu (N_{0,k} + C_{0,k} + I_{k,i})} \right]^+.
\label{eq:power_sense}
\end{equation}
%%%%%%%%%%%%%%%%%%%%%%%
If \( P_{S,k,i} = 0 \), then $\gamma_{k,i} = \frac{(1 + \eta_{k,i}) \lambda_{k,i}}{N_{0,k} + C_{0,k} + I_{k,i}} - \mu.$
\par As for JRC \acp{UE}, the objective is maximizing \( U_k \). The derivative is defined as:
%%%%%%%%%%%%%%%%%%%%%%%
\begin{equation} \small
\begin{aligned}
&\frac{\partial \mathcal{L}}{\partial P_{JRC,k,i}} = \alpha_{k,i} \beta_{k,i} \frac{B}{\ln 2} \frac{\lambda_{k,i}}{N_{0,k} + I_{k,i} + P_{JRC,k,i} \lambda_{k,i}} + \nu_{k,i} \beta_{k,i}\\& \frac{B}{\ln 2} \frac{\lambda_{k,i}}{N_{0,k} + I_{k,i} + P_{JRC,k,i} \lambda_{k,i}} + (1 - \alpha_{k,i}) \beta_{k,i} \frac{\lambda_{k,i}}{N_{0,k} + C_{0,k} + I_{k,i}} \\
&+ \eta_{k,i} \beta_{k,i} \frac{\lambda_{k,i}}{N_{0,k} + C_{0,k} + I_{k,i}}  - \mu \beta_{k,i} + \gamma_{k,i} \beta_{k,i} = 0.
\end{aligned}
\label{eq:kkt_jrc}
\end{equation}
%%%%%%%%%%%%%%%%%%%%%%%
If $P_{JRC,k,i}$$>$0 (\( \gamma_{k,i} = 0 \)), then the \ac{PA} to the \ac{JRC} \acp{UE} is as follows:
%%%%%%%%%%%%%%%%%%%%%%%
\begin{equation} \small
\begin{aligned}
&P_{JRC,k,i} = \left[ \frac{B (\alpha_{k,i} + \nu_{k,i})}{\mu \ln 2} - \frac{N_{0,k} + I_{k,i}}{\lambda_{k,i}} \right. \\& \left.+ \frac{(1 - \alpha_{k,i} + \eta_{k,i}) (N_{0,k} + C_{0,k} + I_{k,i})}{\mu \lambda_{k,i}} \right]^+.
\end{aligned}
\label{eq:power_jrc}
\end{equation}
%%%%%%%%%%%%%%%%%%%%%%%
If \( P_{JRC,k,i} = 0 \), then $\gamma_{k,i} =  \frac{\alpha_{k,i}B}{\ln 2} \frac{\lambda_{k,i}}{N_{0,k} + I_{k,i}} + \frac{\nu_{k,i}B}{\ln 2} \frac{\lambda_{k,i}}{N_{0,k} + I_{k,i}} $ $+ (1 - \alpha_{k,i} + \eta_{k,i}) \frac{\lambda_{k,i}}{N_{0,k} + C_{0,k} + I_{k,i}} - \mu.$
%%%%%%%%%%%%%%%%%%%%%%%%%%%%%%%%%%%%%%%%%%%%%%%%%%
\vspace{-2mm}
\subsection{Solution for Lagrange Multipliers}
As shown in Algorithm \ref{alg:WF}, after the \( P_{k,i} \) values are computed using \eqref{eq:power_comm}, \eqref{eq:power_sense}, and \eqref{eq:power_jrc} for each \ac{UE} type, Lagrange multipliers \( \mu \), \( \nu_{k,i} \), and \( \eta_{k,i} \) are calculated. To begin with, for each \ac{UE} type, \( \mu \) is computed to satisfy the power constraint of \eqref{main} using a bisection search. Let \( \mu_{\text{low}} \) and \( \mu_{\text{high}} \) be initial bounds selected on the feasible power range. The bisection search iteratively updates \( \mu \) as the average of the current bounds, which is defined as $\mu = \frac{\mu_{\text{low}} + \mu_{\text{high}}}{2}.$ The total power is then evaluated using the current value of \( \mu \). If the total power exceeds \( P_{\text{total}} \), the lower bound \( \mu_{\text{low}} \) is set to \( \mu \); otherwise, the upper bound \( \mu_{\text{high}} \) is set to \( \mu \). This process continues until convergence within a predefined tolerance \( \epsilon_\mu \).
\par For QoS constraints, if $C_k < C_{k,i,\text{min}}$ or $\text{SCNR}_k < S_{k,i, \text{min}}$, \( \nu_{k,i} \) or \( \eta_{k,i} \) are adjusted via subgradient updates as follows:
%%%%%%%%%%%%%%%%%%%%%%%
\begin{equation}\small
\begin{aligned}
& \quad \nu_{k,i}^{(t+1)} = \left[ \nu_{k,i}^{(t)} + \delta (C_{k,i, \text{min}} - C_k \beta_{k,i}) \right]^+, \\&\quad \eta_{k,i}^{(t+1)} = \left[ \eta_{k,i}^{(t)} + \delta (S_{k,i, \text{min}} - \text{SCNR}_k \beta_{k,i}) \right]^+,
\end{aligned}
\label{eq:subgradient_update}
\end{equation}
%%%%%%%%%%%%%%%%%%%%%%%
where \( \delta > 0\) is a step size, and \( t \) is an iteration index. This ensures feasible multipliers, satisfying all constraints.
%%%%%%%%%%%%%%%%%%%%%%%
\vspace{-2mm}
\subsection{The Convergence}
\par The above algorithm employs successive convex approximation to converge to a local optimum. Let $f(\mathbf{P})$ denote the objective function. With each \emph{t-th} iteration, a concave surrogate \( \tilde{f}(\mathbf{P} | \mathbf{I}^{(t)}) \) is maximized with fixed interference matrix \( \mathbf{I}^{(t)} \), yielding \( \mathbf{P}^{(t+1)} \) such that \( \tilde{f}(\mathbf{P}^{(t+1)} | \mathbf{I}^{(t)}) \geq \tilde{f}(\mathbf{P}^{(t)} | \mathbf{I}^{(t)}) \). Since interference updates form a fixed-point iteration, \( f(\mathbf{P}^{(t)}) \) converges under bounded channel gains \cite{razaviyayn2013unified}. The process repeats until convergence is reached. We define convergence as the difference $\| \mathbf{P}^{(t)} - \mathbf{P}^{(t-1)} \|_2 < \epsilon$, where $\epsilon$ is a pre-defined threshold. The output is $\mathbf{P}=\{P_{k,i}\}~{\forall k, i}$ for the next $\tau_c$.
%%%%%%%%%%%%%%%%%%%%%%%%%%%%%%%%%%
\begin{figure*}[ht]
\centering      
\subfloat[]{
  \includegraphics[width=55mm]{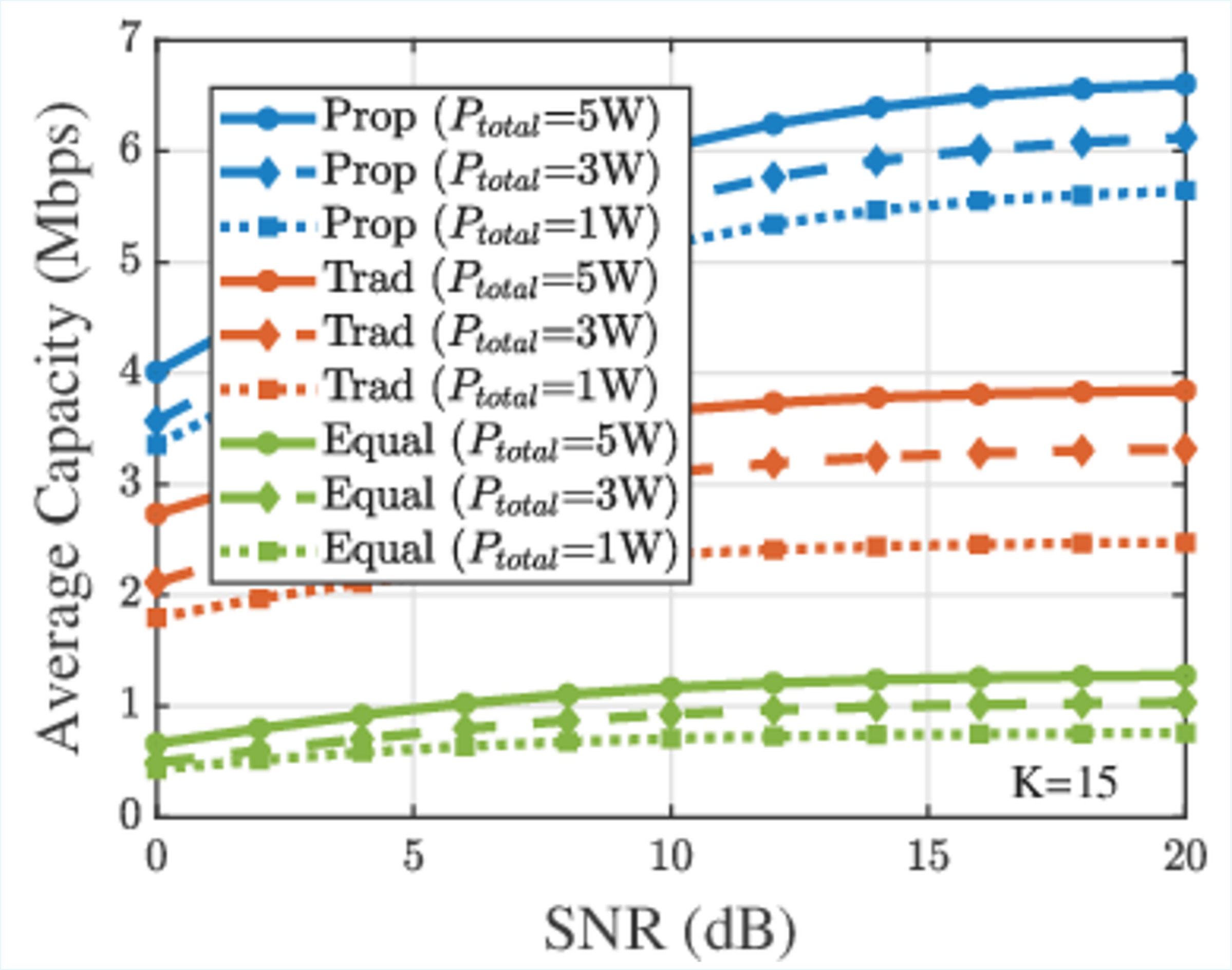}
}
\subfloat[]{
  \includegraphics[width=55mm]{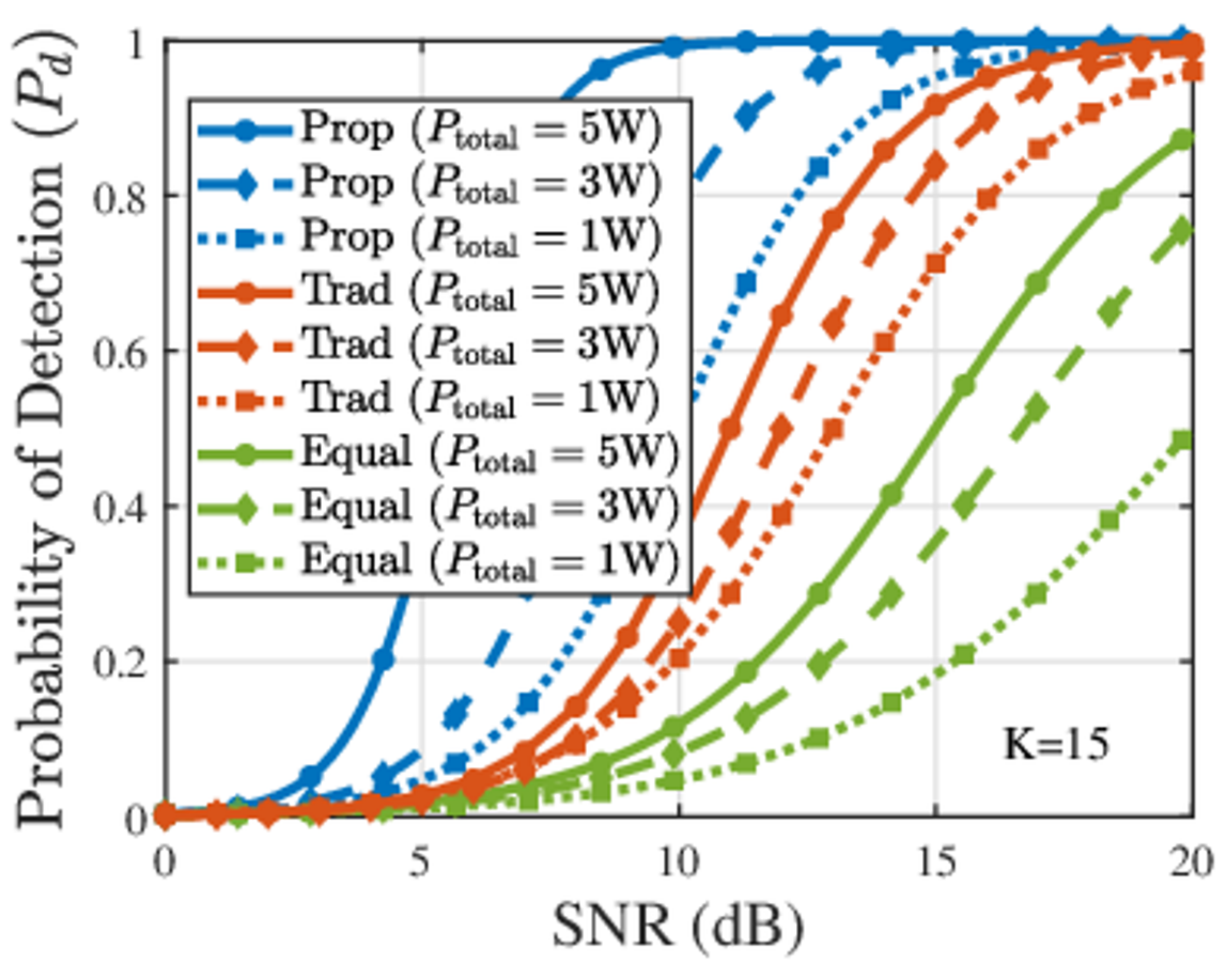}
}
\subfloat[]{
  \includegraphics[width=55mm]{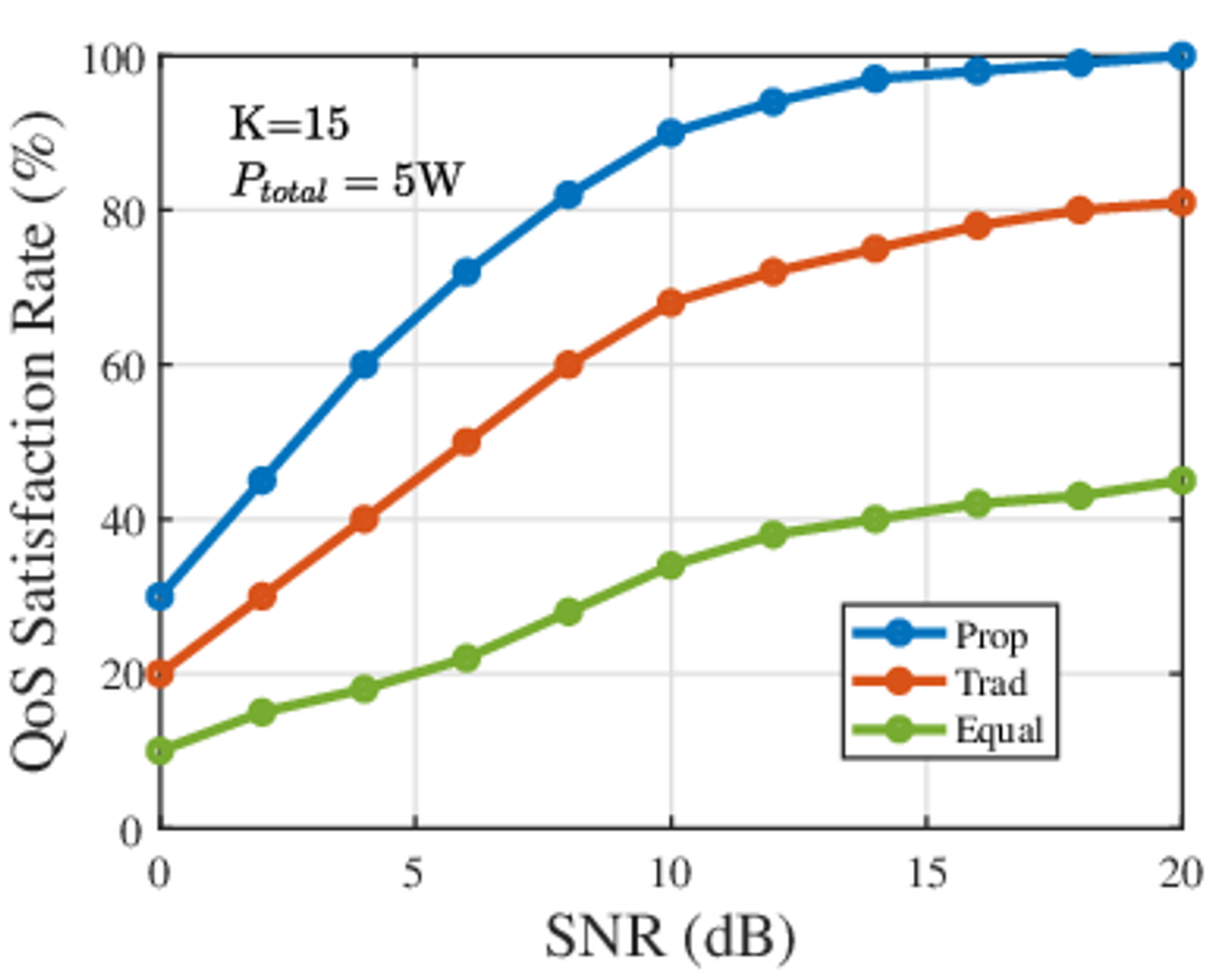}
}
\caption{Comparison between the proposed iterative unified \ac{WF} (i.e., Prop), traditional \ac{WF} (i.e., Trad), and equal \ac{PA} \ac{WF} (i.e., Equal) algorithms: (a) average capacity vs. \ac{SNR} with different values of $P_\text{total}$, (b) the probability of detection vs. \ac{SNR} with different values of $P_\text{total}$, and (c) \ac{QoS} satisfaction rate vs. SNR.}
\label{capa}
\end{figure*}
%%%%%%%%%%%%%%%%%%%%%%%%%%%%%%%%%%%%%%%%%%%%%%%%%%
\vspace{-2mm}
\subsection{Complexity Analysis}
The complexity of the proposed iterative unified \ac{WF} algorithm depends on the number of \acp{UE} $K$, the total number of sub-channels $N = \sum_{k=1}^K r_k$, and the number of iterations $T$ to converge. Per an iteration, the algorithm first performs an interference update. For each $(k,i)$ pair, computing interference has $O(N)$ complexity, resulting in $O(K \cdot \bar{r} \cdot N)$ per iteration, where $\bar{r}$ is the average rank. Furthermore, updating $P_{k,i}$ and solving for $\mu$ using bisection over $N$ variables yields $O(N \log(1/\epsilon_\mu))$. Sub-gradient updates for dual variables \( \nu_{k,i} \) and \( \eta_{k,i} \) add another $O(N)$. Additionally, the convergence check (i.e., $\|\mathbf{P}^{(t+1)} - \mathbf{P}^{(t)}\|_2$) requires $O(N)$. Assuming uniform rank $r_k = r$, we get $N \approx K r$ and $\bar{r} \approx r$. Thus, the per-iteration complexity can be simplified as $O(K^2 r^2 + K r \log(1/\epsilon_\mu))$, and the total complexity can be approximated to $ O\left(T \cdot (K^2 r^2 + K r \log(1/\epsilon_\mu))\right).$ Empirically, convergence is achieved within $T \leq 50$ iterations, yielding a practical complexity of $O(K^2 r^2)$. In contrast, equal \ac{PA} assigns $P_{k,i} = P_{\text{total}}/N$ uniformly, requiring only $O(K)$ operations with no optimization or QoS consideration. However, it fails to adapt to heterogeneous \ac{UE} needs, resulting in poor capacity and SCNR. While the proposed unified iterative \ac{WF} is more computationally intensive, it remains polynomial in terms of $K$ and $r$, which makes it acceptable for practical systems.
%%%%%%%%%%%%%%%%%%%%%%%%%%%%%%%%%%%%%%%%%
\begin{figure}
\centering 
\resizebox{0.55\columnwidth}{!}{
\includegraphics{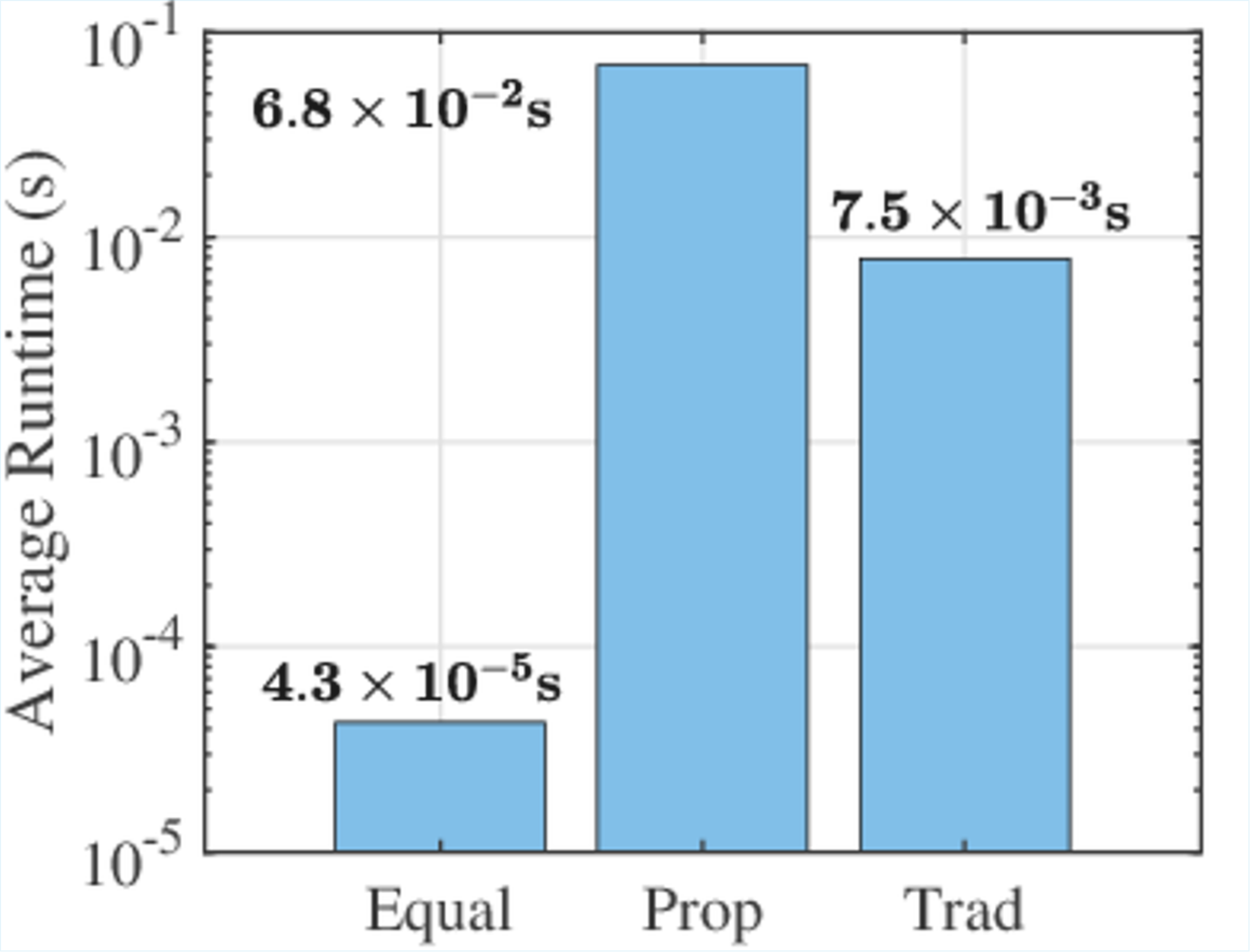}}
\caption{Average runtime for equal \ac{PA}, proposed \ac{WF} and traditional \ac{WF}.}
\label{Figr}
\end{figure}
%%%%%%%%%%%%%%%%%%%%%%%%%%%%%%%%%%%%%%%%%
\vspace{-2mm}
\section{Simulation Results} 
\label{sec:simulation_results}
\par In this section, we present the results for three different \ac{WF} algorithms, the proposed iterative unified \ac{WF}, the existing traditional \ac{WF}, and the existing equal \ac{PA} \ac{WF} \cite{yu2007multiuser}. While in equal \ac{PA} \ac{WF} \acp{UE} are allocated an equal amount of power, traditional \ac{WF} takes into account the communication channel conditions. In our simulations, we assume $K=15$ with $|\mathcal{K}_C|=|\mathcal{K}_S|=|\mathcal{K}_{JRC}|=5$, $N_t=8$, and $N_{r,k}=2$.
%%%%%%%%%%%%%%%%%%%%%%%%%%%%%%%%%%%%%%%%
\par Figure~\ref{capa}(a) shows the average capacity vs. \ac{SNR} for varying $P_\text{total} \in \{1, 3, 5\}$ W. As $P_\text{total}$ increases, the capacity improves due to higher overall transmit power. The proposed algorithm consistently outperforms traditional and equal \ac{PA} \ac{WF} algorithms by adaptively distributing power to satisfy heterogeneous QoS requirements. For example, it outperforms traditional and equal \ac{PA} \ac{WF} algorithms by $2.7$~Mbps and $5.5$~Mbps for SNR=20 dB, respectively. The performance gap is especially evident at low power budgets and becomes more pronounced with increasing \ac{SNR}, though it tends to narrow at very high SNR due to \ac{MU} interference. Nevertheless, in contrast to traditional and equal \ac{PA} \ac{WF}, which saturate at around $14$ dB and $10$ dB of \ac{SNR}, respectively, the proposed algorithm starts saturating only at around $19$ dB. 
%%%%%%%%%%%%%%%%%%%%%%%%%%%%%%%%%%%%%%%%%
\par Figure~\ref{capa}(b) illustrates the \ac{$P_d$} vs. \ac{SNR} for all three \ac{WF} algorithms under varying total power levels, \(P_{\text{total}} = \{1, 3, 5\}\) W. The proposed method achieves superior detection performance across all power levels, reaching \(P_d \approx 1\) more rapidly due to its SCNR-aware optimization that adaptively allocates power to all \acp{UE}. For example, \(P_d \approx 1\) is reached at \ac{SNR}=10 dB for $P_{total}=5$ W. In contrast, the traditional WF and equal \ac{PA} \ac{WF} algorithms suffer from slower growth in detection probability, especially under low power budgets. Notably, the equal power method performs the worst due to its inability to prioritize high-SCNR links. The performance gap between schemes becomes more pronounced as \(P_{\text{total}}\) decreases, highlighting the robustness of the proposed approach in resource-limited scenarios.
%%%%%%%%%%%%%%%%%%%%%%%%%%%%%%%%%%%%%%%
\par Figure \ref{capa}(c) shows the QoS satisfaction rate vs. SNR comparing all three \ac{WF} algorithms for $P_{total}=5$ W. We assume that the \ac{QoS} is satisfied when \(C_{k} \geq C_{k,i,\text{min}}\) and \(S_{k} \geq S_{k,i,\text{min}}\). The proposed \ac{WF} algorithm achieves the highest satisfaction rates across all SNRs, surpassing 90\% at high SNR. The traditional \ac{WF} improves gradually but lags, while the equal \ac{PA} \ac{WF} performs the worst due to its inability to adapt to heterogeneous \ac{QoS} demands. This highlights the effectiveness of the proposed approach in fulfilling diverse \ac{UE}~needs.
%%%%%%%%%%%%%%%%%%%%%%%%%%%%%%%%%%%%%%%%
\par Figure \ref{Figr} compares average runtime for the three algorithms with $P_{total}=5$ W. Although the proposed iterative unified \ac{WF} algorithm incurs the highest runtime due to its iterative nature and adaptive \ac{QoS} considerations, it results in a difference of 0.0605s compared to the traditional \ac{WF} algorithm. The traditional \ac{WF}, omitting \ac{QoS} constraints, achieves lower complexity per iteration. Equal power, with no optimization or interference handling, has the lowest runtime. Despite higher runtime, the proposed \ac{WF} delivers superior performance and fairness among heterogeneous \acp{UE}.
%%%%%%%%%%%%%%%%%%%%%%%%%%%%%%%%%%%%%%%%
\section{Conclusion}
This work proposed a \ac{DL}, interference-aware, QoS-constrained \ac{WF} algorithm, called unified iterative \ac{WF}, for heterogeneous \acp{UE}. By integrating \ac{UE}-specific utility functions, service priorities, and fairness mechanisms, the algorithm dynamically adapts to diverse communication, sensing, and JRC requirements. Simulation results confirm that the proposed algorithm significantly outperforms traditional and equal \ac{PA} methods in terms of capacity, detection performance, and fairness, while maintaining manageable computational complexity. Future work will include addressing the challenge to improve the average runtime.
%%%%%%%%%%%%%%%%%%%%%%%%%%%%%%%%%%%%%%%%%%%%%%%%%%%%%%



\begin{thebibliography}{10}
\providecommand{\url}[1]{#1}
\csname url@samestyle\endcsname
\providecommand{\newblock}{\relax}
\providecommand{\bibinfo}[2]{#2}
\providecommand{\BIBentrySTDinterwordspacing}{\spaceskip=0pt\relax}
\providecommand{\BIBentryALTinterwordstretchfactor}{4}
\providecommand{\BIBentryALTinterwordspacing}{\spaceskip=\fontdimen2\font plus
\BIBentryALTinterwordstretchfactor\fontdimen3\font minus \fontdimen4\font\relax}
\providecommand{\BIBforeignlanguage}[2]{{%
\expandafter\ifx\csname l@#1\endcsname\relax
\typeout{** WARNING: IEEEtran.bst: No hyphenation pattern has been}%
\typeout{** loaded for the language `#1'. Using the pattern for}%
\typeout{** the default language instead.}%
\else
\language=\csname l@#1\endcsname
\fi
#2}}
\providecommand{\BIBdecl}{\relax}
\BIBdecl

\bibitem{10458884}
A.~Naeem \emph{et~al.}, ``Polarization-based multiplexing: Enabling spectrum efficient joint radar and communication,'' \emph{IEEE Wireless Communications Letters}, vol.~13, no.~5, pp. 1414--1418, 2024.

\bibitem{9990579}
A.~Gharib and M.~Ibnkahla, ``Heterogeneous cluster-based information-centric sensor networks with user security satisfaction,'' \emph{IEEE Internet of Things Journal}, vol.~10, no.~9, pp. 8123--8139, 2023.

\bibitem{10436227}
M.~A.~C. Castro and J.~L.~A. Cusicuna, ``Implementation of water-filling algorithm for optimal power allocation in the wireless channel of a {MIMO} 2x2 system over {SDR},'' in \emph{IEEE Colombian Caribbean Conference}, 2023, pp. 1--6.

\bibitem{10989634}
N.~Balasuriya, A.~Mezghani, and E.~Hossain, ``Physically-consistent multi-band massive {MIMO} systems: A radio resource management model,'' \emph{IEEE Transactions on Wireless Communications}, pp. 1--1, 2025.

\bibitem{10908220}
Y.~Dong \emph{et~al.}, ``Resource allocation in multi-user {MIMO-OFDMA} {VLC} systems over low-pass {LED} channels,'' \emph{IEEE Photonics Technology Letters}, vol.~37, no.~11, pp. 633--636, 2025.

\bibitem{yu2007multiuser}
W.~Yu, ``Multiuser water-filling in the presence of crosstalk,'' in \emph{2007 Information Theory and Applications Workshop}, 2007, pp. 414--420.

\bibitem{popescu2007simultaneous}
O.~Popescu, D.~C. Popescu, and C.~Rose, ``Simultaneous water filling in mutually interfering systems,'' \emph{IEEE Transactions on Wireless Communications}, vol.~6, no.~3, pp. 1102--1113, 2007.

\bibitem{viswanath2002asymptotically}
P.~Viswanath, D.~N.~C. Tse, and V.~Anantharam, ``Asymptotically optimal water-filling in vector multiple-access channels,'' \emph{IEEE Transactions on Information Theory}, vol.~47, no.~1, pp. 241--267, 2002.

\bibitem{wang2015iterative}
Z.~Wang, V.~Aggarwal, and X.~Wang, ``Iterative dynamic water-filling for fading multiple-access channels with energy harvesting,'' \emph{IEEE Journal on Selected Areas in Communications}, vol.~33, no.~3, pp. 382--395, 2015.

\bibitem{naser2023downlink}
M.~A. Naser \emph{et~al.}, ``Downlink training sequence design based on waterfilling solution for low-latency {FDD} massive {MIMO} communications systems,'' \emph{Electronics}, vol.~12, no.~11, p. 2494, 2023.

\bibitem{chauhan2023power}
A.~Chauhan, S.~Singh, and P.~Gupta, ``Power optimization for millimeter wave {MIMO} system,'' in \emph{International Conference on Computing Science, Communication and Security}.\hskip 1em plus 0.5em minus 0.4em\relax Springer, 2023, pp. 233--244.

\bibitem{razaviyayn2013unified}
M.~Razaviyayn \emph{et~al.}, ``A unified convergence analysis of block successive minimization methods for nonsmooth optimization,'' \emph{SIAM Journal on Optimization}, vol.~23, no.~2, pp. 1126--1153, 2013.

\end{thebibliography}
\end{document}